# A New Fuzzy MCDM Framework to Evaluate E-Government Security Strategy

Irfan Syamsuddin [1,2] Junseok Hwang [1]
[1] *International IT Policy Program*
*College of Engineering, Seoul National University,*
*Republic of Korea*
[2] *State Polytechnic of Ujung Pandang,*
*Republic of Indonesia*
irfans@poliupg.ac.id, junhwang@snu.ac.kr

*Abstract* - Ensuring security of e-government applications and infrastructures is crucial to maintain trust among stakeholders to store, process and exchange information over the e-government systems. Due to dynamic and continuous threats on e-government information security, policy makers need to perform evaluation on existing information security strategy as to deliver trusted e-government services. This paper presents an information security evaluation framework based on new fuzzy multi criteria decision making (MCDM) to help policy makers conduct comprehensive assessment of e-government security strategy.

## I. INTRODUCTION

E-government is about bridging government and citizen communications in more efficient, transparent and reliable ways through effective use of information technology.

The Internet has become the main media for e-government from delivering public information to electronic document and financial transactions although it is widely attributed to serious security weaknesses. As a result, security and privacy are the most crucial concerns of any e-government applications [1].

Many attempts have been proposed to overcome the issue. Unfortunately, most of the previous solutions tend to focus on technical solutions in various forms details such as firewall, intrusion detection systems, web application security and penetration testing using proprietary or open source technologies. In fact, security is no longer solely a technical concern; it is now becoming complex issues [2, 3].

Lack of adequate framework which combines broad perspectives of e-government is considered as a gap in recent literature [3, 4] particularly when dealing with how to evaluate information security practices of e-government for changing strategy in the future.

Ensuring security of e-government applications and infrastructures is crucial to maintain trust among stakeholders to store, process and exchange information over the e-government systems. Due to dynamic and continuous threats on e-government information security, policy makers need to perform continuous evaluation on existing information security practices and controls. Based on the fact, this paper attempts to propose a holistic approach from managerial decision making perspective by combining all related aspects of security to create a framework used to evaluate e-government security strategy.

Since decision making mostly involve multi criteria and alternative to consider altogether, this framework implement multi criteria decision making (MCDM) approach to view e-government security strategy from managerial perspective.

Fuzzy set theory is applied to complement the framework in order to capture fuzziness in the form of inconsistencies and vagueness coming from subjective judgments by decision makers.

The rest of this paper is organized as follows. Section 2 presents literature review of e-government security. In section 3, basic concept of fuzzy set theory is described. Then, in section 4, the new fuzzy multi criteria decision making framework is clearly explained and justified. Finally, concluding remarks and future research directions are given in last section.

## II. E-GOVERNMENT SECURITY

E-government is one of key indicator for nations development as formulated by United Nation Public

Administration Networks (UNPAN) [6]. Proliferation of e-government strongly relies on trust among citizens to store, process and exchange information over the e-government systems. Trust is maintained through effective security controls to ensure no sensitive information goes to unauthorized person.

In some studies [7, 8] security issues are found to affect public services management. Trust on e-government plays a significant role to improve efficiency and effectiveness on transparent information flow between governments business and citizens [9], On the other hand, lack of security concerns are responsible for unsuccessful e-government initiatives in some developing countries [10]. Therefore, security controls is one of key factors for achieving an advanced stage of e-government for national development.

Whilst technical considerations have received major concern in improving security controls on e-government systems in the past [11], non-technical issues such management, economic and cultural issues have emerged since the impact of cyber crimes has also become widely affecting many government and business organizations [2,12,13]. Complying with international standards, strong management commitment and regular review of security controls for security updates are examples of management aspects of information security that has received increasing concerns in recent years [13].

E-government security might also be viewed from economic perspectives. Security related spending with adequate investment analysis is believed to bring more success in handling security threats [12] which critically measure risks associated with impacts of security threats on e-government sensitive information. Van Solms [13], on the other hand, argues that success in maintaining information security mainly depends on people. Security awareness as a means of human security can be improved through comprehensive education and implementing reward punishment mechanisms [13]. This creates the so called security culture. Gap of security between members and management of the organization about information security should be reduced otherwise it potentially become another source of threats in the future. Therefore in order for e-government to be implemented properly, security culture should be taken into top list of consideration.

While the vital role of information security is significantly justified, only limited studies discuss specifically on how to evaluate and to make decisions regarding e-government security strategy.

Although different organizations tend to define different security strategies, they generally agree upon the basic security strategy based on the CIA security triangle (confidentiality, integrity, and availability [14]. Strategic decision lies on fundamental question of how to allocate security resources within the three security elements.

One of a significant contribution to answer this question was proposed by Hwang and Syamsuddin [3] who introduce an information security policy decision making based on Analytic Hierarchy Process. This work was expanded and applied to e-banking security to guide policy makers in performing evaluation of e-banking security in Indonesia [5]. However, this approach did not tolerate fuzziness such as inconsistency and vague decisions which addressed in many MCDM literatures. For that reason, in the following sections, we describe how fuzzy set theory is adopted and combined with MCDM to develop a new fuzzy multi criteria decision making for e-government security purposes.

## III. FUZZY SET THEORY

The fuzzy set theory was introduced by Zadeh [15] to deal with fuzziness issues in many control systems applications. It was oriented to the rationality of uncertainty due to imprecision or vagueness. Its ability in representing vague data is considered as the major contribution of fuzzy set theory to science and technology. In the area of MCDM, fuzzy set theory has given a significant contribution by accepting uncertainty and inconsistent judgment as a nature of human decision making [16,18,19].

In fuzzy set theory, triangular fuzzy numbers are represented with a triplet (L, M, U) for Lower, Medium and Upper numbers. Figure 3 shows the membership triangular fuzzy numbers.

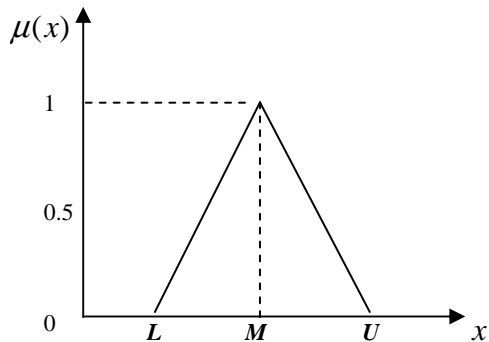

Figure 1. Fuzzy triangulars membership function

Let A be a triangular fuzzy number with a triplet (L, M, U). The membership value can be defined as follows

$$u_A(x) = \begin{cases} \dfrac{x-L}{M-L}, & L \le x \le M \\ \dfrac{x-L}{M-L}, & L \le x \le M \\ 0, & \text{otherwise} \end{cases}$$

Various arithmetic calculations can be applied to these fuzzy membership values. However, there are few basic operations that widely used for triangular fuzzy numbers as explained below.

Let $\tilde{M} = (L_1, M_1, U_1)$ and $\tilde{N} = (L_2, M_2, U_2)$

Thus, fuzzy number reciprocal:

$$\tilde{M}^{-1} = (1/U_1, 1/M_1, 1/L_1)$$

and

$$\tilde{N}^{-1} = (1/U_2, 1/M_2, 1/L_2) \quad (1)$$

Fuzzy number addition:
$$\tilde{M} \oplus \tilde{N} = (L_1+L_2, M_1+M_2, U_1+U_2) \quad (2)$$

Fuzzy number subtraction:
$$\tilde{M} - \tilde{N} = (L_1-U_2, M_1-M_2, U_1-L_2) \quad (3)$$

Fuzzy number multiplication:
$$\tilde{M} \otimes \tilde{N} = (L_1.L_2, M_1.M_2, U_1.U_2) \quad (4)$$

Fuzzy number division:
$$\tilde{M} \div \tilde{N} = \left(\dfrac{L_1}{U_2}, \dfrac{M_1}{M_2}, \dfrac{U_1}{L_2}\right) \quad (5)$$

A major contribution of fuzzy set theory is its capability of representing vague data. The theory also allows mathematical operators and programming to apply to the fuzzy domain [15].

In this section we introduce the proposed method as the main objective of this study. The new fuzzy multi criteria decision making framework for e-government security strategy is aimed at providing comprehensive decision making solution with ability to deal with inconsistent and vague judgments during decision making processes by the policy makers.

Instead of using crisp numbers to represent preference used in classical Analytic Hierarchy Process [17], fuzzy numbers along with its linguistic variables are applied in this framework as shown in the following table.

Table 1. Fuzzy scales

| Linguistic Variable | Fuzzy Scale (*l.m.u*) | Reciprocal scale (*l.m.u*) |
|---|---|---|
| Equally Important | (0.5,0.5,0.55) | (0.45,0.5,0.5) |
| Slightly Important | (0.55,0.6,0.65) | (0.35,0.4,0.45) |
| Important | (0.65,0.7,0.75) | (0.25,0.3,0.35) |
| Very Important | (0.75,0.8,0.85) | (0.15,0.2,0.25) |
| Absolutely Important | (0.85.0.9,0.9) | (0.1,0.1,0.15) |

Comprehensive analysis of the framework is discussed in the following steps:

### A. Construction of hierarchy

The proposed framework is basically adopted from the Analytic Hierarchy Process to form the hierarchy. As shown in figure 2 it consists of four levels, goal, criteria, sub criteria, and alternatives.

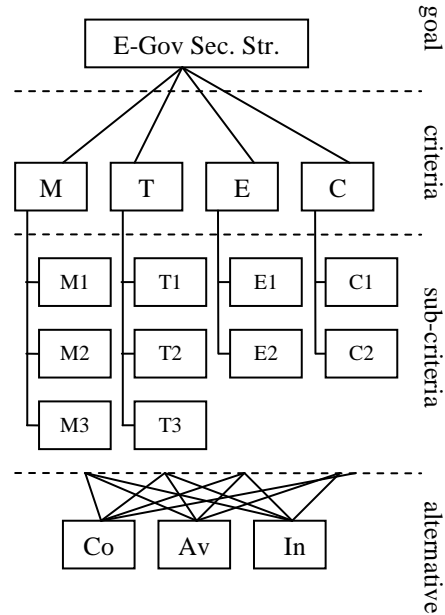

Figure 2. New fuzzy MCDM framework for E-government security strategy.

At the top level, we define the goal as information security policy performance evaluation. Subsequently, four criteria are listed at the second level, namely

Management (M), Technology (T), Economy (E) and Culture (C) as the criteria.

At the third level, all criteria are divided into sub criteria as follows. Management consists of comply with standard (M1), regular review (M2), and commitment (M3). Technology consists of end point security (T1), network security (T2) and application security (T3). Economy consists of security investment (E1) and cost of attack (E2). Culture consists of reward & punishment (C1) and security education (C2).

Finally, at the bottom level, we define three security objectives as the central concern in making any security decisions. They are confidentiality integrity and availability or commonly called CIA Security Triangle [3,5,14].

### B. Pairwise Comparison by Decision Makers

Based on the analytic hierarchy process structure [17], a set of questionnaire should be formed based on linguistic variables mentioned in table 1.

The decision maker will be headed with a series of pairwise comparison questions in the the general form such as, 'How important is criterion M relative to criterion E ?'. The optional answer is available six linguistic variable of *Just Equal, Slightly Important, Important, Strongly Important, Very Strongly Important*, and *Absolutely Important* (see table 1)

Later all answers are translated into its corresponding fuzzy scale as well the reciprocal numbers.

### C. Aggregating Fuzzy Weights

In this framework we adopt fuzzy geometric mean method [16, 18] to deal with N numbers of decision makers since it has specific characteristic in shortening the gap effects of very high or low values, which significantly affect in reducing estimation bias.

Calculate geometric mean first for "upper" fuzzy numbers.

$$\tilde{g}_u = (\tilde{g}_{u1} \otimes \tilde{g}_{u2} \otimes \tilde{g}_{u3} \otimes .... \otimes \tilde{g}_{un})^{1/n} \quad (6)$$

The weight for upper fuzzy numbers derived from normalizing geometric mean

$$\tilde{w}_u = \frac{\tilde{g}_u}{\sum_{1}^{n} \tilde{g}_{un}} \quad (7)$$

Later, we will obtain $\tilde{w}_l; \tilde{w}_m; \tilde{w}_u$ as the fuzzy numbers for lower, medium and upper.

### C. Calculating final weight

Overall weight is calculated exactly the same with classical AHP method [17]. In this case, all fuzzy numbers derived at each level are constructed in the form of matrix and then multiplying them to derive final weight. This step is conducted onto three fuzzy numbers (*l,m,u*).

### D. Aggregating all decision makers final weights

Since several decision makers (N) are involved, the final analysis step of the framework is to aggregate all of these decisions. The calculation is performed through arithmetic mean to derive final weight for all decision makers as follows

$$\tilde{w}_{fl} = \frac{\sum_{1}^{n} w_{ln}}{n} \;;\; \tilde{w}_{fm} = \frac{\sum_{1}^{n} w_{mn}}{n} \;;\; \tilde{w}_{fu} = \frac{\sum_{1}^{n} w_{un}}{n} \quad (8)$$

Finally, we obtain three final weights above for lower, medium and upper fuzzy numbers. The three values are used to describe pessimistic, normal, and optimistic modes for further simulation purposes.

In this framework, the application of fuzzy set theory on AHP by Buckley [15] is modified and improved to meet the requirement for developing a new fuzzy MCDM framework to evaluate e-government security strategy.

## IV. CONCLUSION

This paper introduces a new fuzzy multi criteria decision making framework for e-government security strategy. Its main feature lies in its ability to capture vagueness and inconsistencies coming from subjective human judgments as decision makers. This feature overcomes weaknesses on previous approaches based on classical AHP.

In details, comprehensive calculation procedures of the fuzzy MCDM framework have been clearly explained and justified which can be implemented in real case. It starts from gathering decision makers' judgments with fuzzy numbers and finally aggregates their fuzzy judgments into three modes of analysis, pessimistic, normal and optimistic.

Furthermore, we plan to extend this study by applying it into empirical case study to assist policy makers in making appropriate decisions under fuzzy situation.